# Business Value of ITSM – Requirement or Mirage?

**James J. Cusick, PMP**

IEEE Computer Society Member

New York, NY

*j.cusick@computer.org*

**Abstract—** *This paper builds on a presentation provided as part of a recent panel session on ITSM (IT Service Management) Business Value at the NYC itSMF (Service Management Forum) Local Interest Group meeting. The panel presentation explored the definition of Business Value and how ITSM itself could be measured to produce business value. While ITSM and ITIL have been in use for years it often remains a challenge to demonstrate the business value of these methods or even to understand business value itself. This paper expands on the panel discussion on what is meant by business value and how it can be found - if at all – in the context of ITSM development and process improvement.*

## I. Introduction

Businesses and corporations have essentially one goal in mind – to make and improve profit[1]. There are many factors that go into making this happen including effective business strategy, clear product strategy, excellent customer service, efficient cost structures, and a work culture attracting and retaining the best talent. For the ITSM (IT Service Management) community the notion of how improved IT Service processes support the business objective of profit realization or "Business Value" has long been a key topic. With the emergence of ITIL 4 the new framework of best practice supporting ITSM implementation Business Value has been placed more squarely at the center [BAS19]. In this paper we explore the definition of Business Value, the difficulty in tracing its impact from ITSM, and share practical suggestions on framing the attainment of Business Value through ITSM.

ITSM refers to all elements of organization, planning, and delivery of IT services. The most popular and widely accepted framework for ITSM and IT Governance has consistently been ITIL (formerly the IT Infrastructure Library) [BON05]. Other popular alternative ITSM models include COBIT, ISO, MOF, Lean, SIAM [HER17] and even approaches like DevOps have taken hold. In this paper the focus is on an ITIL view of ITSM as these both relate to the generation of Business Value as was discussed in the recent panel session [CUS19a] hosted by the itSMF [itSMFa][itSMFb]. This will help ground the discussion in one approach as opposed to attempting a survey of frameworks or a comparison of them.

While the relationship between ITSM and Business Value is often discussed and, in many cases, presumed - it is not always proven or obvious. Thus, this paper will first take a critical look at what is meant by Business Value with respect to ITSM and in general. We will also attempt to explore some fundamental questions in this area including the following:

1. Have ITSM and ITIL contributed towards Business Value Creation?

---

[1] For other forms of corporations and organizations such as non-profits they typically will attempt to maximize their objectives which may lay outside profit seeking. This is a focus outside the specific scope of this paper but the principles here can be applied to many types of organizations.

2. Does Business Value Creation need proof in advance of ITSM implementation, after the fact, or not at all?
3. Is there a specific model and maturity level that can connect ITSM to Business Value needs?

## II. WHY IS DETERMINING BUSINESS VALUE OF ITSM CHALLENGING?

### A. An Accidental Sighting and Business Value

On a recent trip to Japan the author noticed an unusual bird in the suburban hills. The bird was an Izu Island Thrush which is not common on the main island of Japan called Honshu [CUS11]. The bird cannot fly far from its native habitat 500 Kilometers to the South East in the Izu Islands and is in fact endangered [MAS82]. For Birders this is called an "accidental". A bird out of its usual place. Most likely it was pushed off its home islands by a storm and made its way South. Due to this the sighting was unusual and valuable. In fact, because of its rarity in this location the sighting was important.

This event came to mind in preparing to discuss the Business Value of ITSM. For most business leaders ITSM is not well known. Additionally, it is not always common or possible to predict Business Value as driven by ITSM efforts *a priori* or even after the fact. Thus, just like a rare bird which has intrinsic value when sighted, perhaps ITSM has value even if Business Value cannot be determined. To explore this a bit, consider how many process improvements are well-intentioned accidents. Alternatively, many ITSM efforts are pursued more by instinct and experience than by a quantitatively driven approach like Six Sigma. So, perhaps ITSM driven Business Value Creation is just as vulnerable as a rare bird in the wild but also just as precious – **nearly a mirage and not a requirement of planning**.

### B. Doing the Obvious

It is common to discuss Business Value Creation as a process that is well constructed, deliberate, and predicted in advance. However, this level of careful forethought is often not realized. Instead, like the accidental bird, people intuit sources of improved Business Value and create plans and approaches to reach to those enhanced levels.

A strong example along these lines is the work of Taiichi Ohno who helped develop "The Toyota Way" which was later codified as Lean Manufacturing. This approach now dominates not only automobile production but also many other manufacturing sectors [WOM07]. Moreover, Lean has become a critical influence in the Agile Development arena as well.

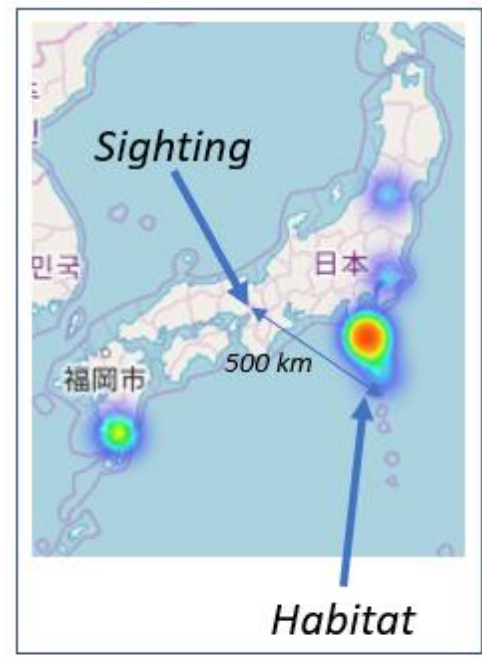

*Figure 1 – An Accidental Sighting*

What is interesting in the context of this paper is that Ohno rarely conducted any formal studies nor did he use elaborate models to design and improve the processes used or introduced. Using several key philosophical principles around respect for people, limiting work in process, small batch sizes, eliminating waste, and focusing on continuous improvement, he relentlessly drove efficiency, yield, and quality. An example is his insight that maintaining high levels of inventory was costly in terms of storage and capital investment and reduced flexibility. This prompted his development of Just-in-Time delivery which produced massive competitive advantage for years. In then end this single-minded approach drove Toyota to be the largest vehicle producer in the world with a fanatical customer base. As for Business Value this approach produced a continuous 50-year increase in market valuation to the current valuation of $214B (see Figure 2 below). This is the true measure of any corporation's success at improving Business Value.

From this philosophy of process improvement – essentially less is more – we see major business value advantages arising over decades. In Ohno's world view, if you see something that is broken or can be improved, do you not simply fix it? This then leads automatically to new value. In effect, following one's gut feeling to pursue the principles of Lean then generates observable value on the backend. This obviates the need for a business case or an extensive Six Sigma forward analysis. An interesting example of this was where Ohno mentions in an interview that he visited a warehouse and told the manager to eliminate it within a year. When he came back in a year it was gone and efficiencies in the supply chain had risen [OHN95]. Again, this was simply based on a principle not on a complex process model.

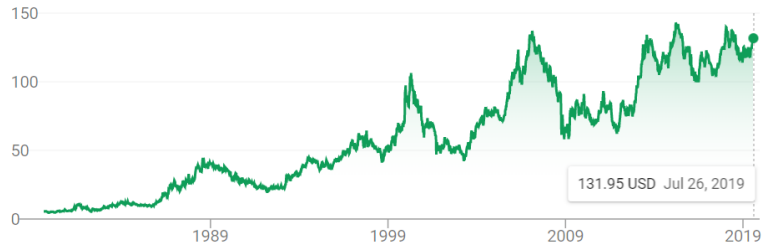

*Figure 2 – NYSE Stock Price in USD of TM (Toyota Motor Company) 1979-2019 (Source Google Finance)*

## C. Invisible Operations Processes

Finally, by way of understanding the challenges of initially determining the value of Business Processes, it is important to recognize that such processes are often embedded deeply in the fabric of the organizations which they serve and in fact can oftentimes be seen as inseparable from them. This leads the observer with a difficult dichotomy of attempting to parse out what is essential to the nature of the business and what is stylistic approach or even "folklore" [KEN97]. These and other factors lead many managers to focus on operations but not necessarily process engineering. As such, continuous process improvement may be less of a focus if operational processes are not explicitly defined and exposed. While these processes do exist implicitly, they are essentially invisible as they are not defined, repeatable, or measurable. When this is the case improvement can happen organically and Business Value can be achieved but proof of that value creation is nearly impossible to determine as there is no framework to determine its scale or velocity. Thus, a process model is required, and a defined process should be in place with appropriate measures and governance to have a chance at achieving improvements in Business Value from ITSM processes.

## III.  HOW DO WE DEFINE BUSINESS VALUE?

### A. Definition of Business Value

Business value is an **informal** term that includes all forms of value that determine the health and well-being of the firm in the long run [WIK19].

> *"'Business value' does not have a single, agreed-upon definition; however, examining several sources that attempt to define what business value is can give us more insight into its meaning and application."* [PHI14]

Business value expands the concept of value beyond economic value (also known as economic profit, economic value added, and shareholder value) to include concepts like employee value, customer value, supplier value, channel partner value, alliance partner value, managerial value, and societal value. Many of these forms of value are not directly measured in monetary terms.

### B. Criticisms

When we look at Business Value itself there are ample issues with the term. First, the term is an "informal" one where there is no consensus in academia or with management professionals as to a standard meaning [WIK19]. This also limits its utility in analysis or decision making. For some, this term can be seen as a "buzz word". In many cases the concept is put forward by consultants, authors, and academics but is not clearly usable by IT practitioners in a clear manner. In the recent iteration of ITIL (ITIL 4) "value creation" or value "cocreation" becomes a central focus even though the broader concept of Business Value remains undefined. The alternative approach might be to rely on traditional economic or business approaches for valuing businesses including market capitalization, profit, and shareholder value. It might be possible to link process improvements under an ITSM framework to true business value through such economic measures, however, this would need to be an intermediate path to workable measurement.

## IV.   CALCULATING THE BUSINESS VALUE OF ITSM

A useful approach to understanding business value is to take a step back from the ITSM perspective and look at business value definition from the broader economic or financial perspective. As such, in finance, business value relates to the value of the firm as derived from traditional calculations such as the:

- Market approach, or;
- Income approach, or;
- Asset approach.

These methods are well documented and have been used extensively [WIK19]. They are considered standard approaches in finance and business applications. However, the use of ITSM within a business might deliver a component of Business Value but typically will not reflect all of Business Value itself. Additionally, most Businesspeople do not care about nor are aware of ITSM. In general, they would also rather not pay for it as they do not put a value on ITSM Business Value itself. It is hard to value something one is not clear on or considers unimportant. This is the essence of valuation in ITIL 4 terms.

What is true is that business stakeholders will care and place a high value on IT Operations and indirectly on ITSM should system Availability become disrupted. Disruptions of service that impact revenue, profit, or customer satisfaction do have a way of getting the attention of the business. Naturally, this is a highly unfortunate situation which both IT and the business strongly work to avoid but it does underscore how a valuation of IT Operations (the daily operations of networks, computers, and the like) and ITSM (the service management and process that control those process) can be delineated. In fact, a major global study of ITIL adoption and the payback of such implementation found that 66% of respondents in the US and UK reported benefits in service quality, 58% in standardize processes, and 48% in improved customer satisfaction [GAC10].

## V.   ITIL AND BUSINESS VALUE

### A.  ITIL Maturity and Business Value

When attempting to determine Business Value a legitimate question revolves around the process and capability maturity of the target organization. For organizations performing at lower maturity levels it may be more challenging to determine increments of value delivered. In traditional maturity models, measurement systems and optimization of processes are found at the higher end of the scale [CUS19b]. This makes it potentially more difficult to define and observe Business Value impacts at the foundational maturity levels where these capabilities are more limited. The typical approach here is to assess the organization, determine the specific improvement areas, and build corrective plans. This may not include qualitative or quantitative methods to up front argue the Business Value benefits or ROI of such process improvement efforts. Instead a bit of a leap of faith must be taken – often by the organization's leadership. This follows in the path set by Ohno. Once the organization has climbed up the maturity ladder sufficiently then quantitative methods can be employed to demonstrate value.

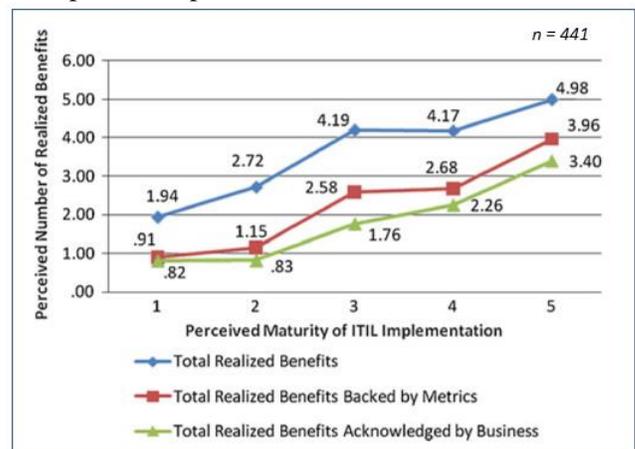

*Figure 3 – Survey of Business Executives on Perceived Value of ITIL vs Measured Benefits Against Maturity Levels [MAR10]*

An extensive study relevant to this discussion was performed by Marrone [MAR10]. In this work, over 400 Business Executives within organizations where ITIL had been implemented were surveyed. In this survey the Business leaders were asked a number of questions around their perceived value of ITIL (essentially synonymous with ITSM or Service Management for the purposes of this discussion). Furthermore, the researchers collected information on measured benefits around the actual processes as used such as improved changes and cycle times, etc. Finally, they plotted all of this data along an axis of assessed ITIL maturity as is shown in Figure 3. At least two clear macro findings emerge from the results. First, the Business Executives consistently underestimate the value of ITIL as compared with the measured value of the ITIL process as per the measurement systems of the process teams. Second, as ITIL maturity levels increase so does the perceived value by the Business leaders. This provides strong reason to invest in higher maturity around ITIL as an ITSM implementation. And to some extent this may make the case for improved communications around IT metrics to bridge the perception gap with management.

## B. Focus on Business Value with ITIL 4

Within the emerging best practice framework of ITL 4 there is a strong emphasis on Business Value Creation. ITIL 4 defines value as: "The perceived benefits, usefulness, and importance of something" [BAS19]. Such value is understood to be derived from the delivery of products and services to or by a set of stakeholders including: Service consumers, Service provider, Service provider employees, Society and community, Shareholders.

The Service Value System - stressed in ITIL 4 (see Figure 4 below)– as a value stream (steps an organization takes to create and deliver products and services to consumers) redefines Business Value for ITSM to be more limited to this view and does not formally encompass the traditional Business Valuation models noted above. Instead "Utility" and "Warranty" drive value outcomes [SMA18] under a set of guiding principles. This means that services do something and do them within given requirements. Warranty refers to "non-functional requirements such as availability, performance, and security". Thus, a service is delivered and is expected to behave conformant to these non-functional requirements meeting these warranties. The transformation of these opportunities into realized values are achieved through what ITIL 4 calls a Service Value Chain. This emerging best practice will serve to inform future discussions of the use of ITIL and ITSM in general for the delivery of Business Value and will provide context during the remainder of this discussion.

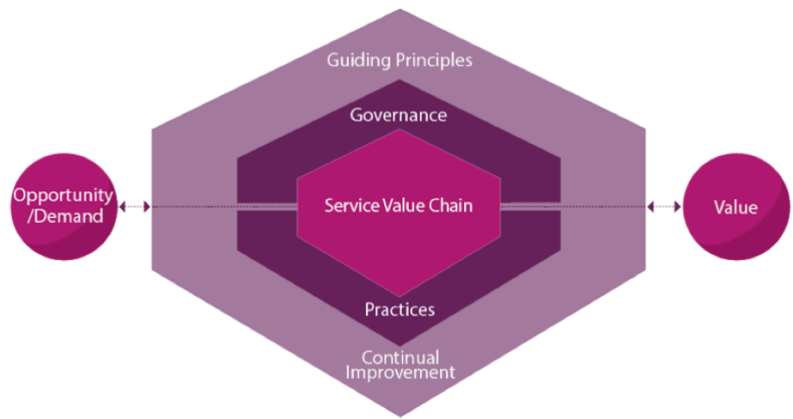

*Figure 4 – The ITIL 4 Value Stream Construction*

## VI. Considerations around Business Value – How to Understand the Challenge

Considering the above, we can consider a few specific questions which can highlight a variety of additional considerations around the question of ITSM and Business Value. To recap so far, we have said that it is not always straightforward, necessary, or even possible to measure ITSM Business Value in practice. However, for ITIL and ITSM we assume they do provide value. So, let us look as some further details.

1. **What are the best ways to measure Business Value?** – As there are literally dozens of potential measurement approaches around Business Value a better question is how to get to your measures of interest. Core Business Value metrics such as top line revenue, revenue growth, bottom line expense, expense management, customer satisfaction, and employee engagement are all standard fare. However, it is possible that by working through a structured measurement definition process for your business with your stakeholders you may find other key metrics which are more relevant to your conditions such as long term quality of life, investment returns, development of talent, brand value, and competitive advantage [SPA18]. One of the best methods for doing this is to first understand what business goals you are trying to achieve. And the simplest and most powerful method to do this is to follow the GQM technique [BAS90]. Goal/Question/Metric is an approach developed by Vic Basili of the University of Maryland to identify metrics of importance in Software Engineering and IT. The method is clear and straightforward. Define what goals of strategic intent need to be answered (such as is your revenue target being met or what your cost reduction objective is). Then define the questions which will inform you that those goals have been met and this will then lead you automatically to the metrics that you need to capture. It is a proven method with a deep research base behind it [BIR10]. Once these measures are identified a classically successful approach to communicating them to an Executive audience is through a dashboard or scorecard such as the Balanced Business Scorecard [KAP96]. This method condenses only the most important metrics in a summarized view for reporting and can be highly effective if used well in conjunction with GQM.

2. **What level of ITSM maturity is required to drive Business Value?** – In a recent historical survey of Maturity Models from their roots in process and quality improvement to the modern day including Agile and DevOps (see CUS19b) a useful context for the application of these models is provided. In reviewing these many different Maturity Models, it becomes clear that we cannot put the cart before the horse. At the same time, it is certainly possible to lay the groundwork for achieving some degree of Business Value even at primitive levels of Maturity. For example, Incident Management is a fundamental IT operational process function, yet it provides significant Business Value by increasing Availability by, among other things, reducing MTTR. Higher level process functions that might be tackled later up the Maturity ladder such as defined SLAs and Capacity Management naturally provide Business Value also but can come in due time.

3. **What process functions have the highest impact on Business Value creation?** – As mentioned above, fundamental process areas such as Incident, Problem, Change, and Release will have tremendous Business Value if they are being provided effectively. However, once the maturity level of the IT organization has climbed higher, selected process areas can then prove much more valuable overall. The foundations become table stakes and the business stakeholders look for more from their IT partners. In recent years the increased focus on Cybersecurity has driven IT to anticipate new solutions and services and get ahead of the problems that might face the business in compliance, tooling, and user training. This type of leadership from IT has provided true Business Value measured in new sales based on customer confidence in security posture. Similar approaches can be taken across other IT Services domains.

4. **How long does it take for implemented processes to show impact on Business Value?** – The lag time for a process to generate returns is somewhat difficult to predict as there are many variables to consider and many types of process changes possible. In the case of an implementation of a ground up ITSM process environment it is likely that a 6-24-month time frame will be required as there needs to be an assessment period, a process development stage, deployment, and then a monitoring and measurement period to quantify results. Of course, in some cases, limited results can be seen much earlier where incremental benefits are built in. In other cases, benefits can take longer to accrue as in one study where reductions in helpdesk tickets began providing benefits early and continued to generate ROI consisting of cost benefits over 3-4 years [SUL13].

5. **What are the best practices or recommendations to implement ITSM?** – A good place to start in thinking about ensuring a successful change and implementation of an ITSM process is the ITIL Organization Change Management practice [HUN16]. These core suggestions include: a) Ensuring strong and committed leadership; b) Helping people understand what the organization is trying to accomplish and why; c) Getting individuals to participate willingly – answering the question "what is in it for me?"; d) Creating prepared participants; and e) Sustaining the momentum by reinforcing the change. This guidance is reinforced by other change management advisors such as the popular Kotter Change Model [KOT12]. Kotter's model is more complex but boils down to preparing a change climate, engaging for change, and sustaining the change. Regardless of the change model, each organization will be unique and find its own challenges. The key is to adapt the change approach to those conditions and to persevere in the face of the inevitable challenges that will be met [CUS18]. In addition, it is important to start small and make progress incrementally. Trying to boil the ocean can sink a well-intentioned project. Finally, always keep in mind that ITSM involves people, process, and technology. Thus, each aspect needs to be attended to from staff roles, training, communications, change management, buy-in, and process design and platform roll-out.

6. **How do tools play a role in the successful implementation, operations, and Business Value Creation with ITSM?** – For ITSM implementation and operations tools play a critical role. Large organizations need to embed their process designs in a dynamic tool environment or substrate in order to make the process visible, support its enforcement, and allow for its continual managed evolution. While any tool will do in theory clearly some tools may be better suited than others to a particular organization and conducting a fact-based due diligence is important when selecting a given ITSM tool base. Once such a tool or tool suite is selected the process standards and designs need to be translated into workflows, procedures, and Knowledge Base references for associates and engineers to follow. This can require a measure of customization and some trial and error which is to be expected. This is also an ongoing process of support for any ITSM automation environment. However, the benefits are significant as manual steps can be reduced to minutes, repeatability built-in, and new levels of quality metrics can be established.

7. **How does the emergence of ITIL 4 play into the future of ITSM adoption, evolution, and especially the ability to demonstrate the achievement of ITSM Business Value?** – To a large extent predicting the future is a fool's errand. However, the trajectory of the ITIL releases have been fairly spaced with several years placed between them. In each case, the releases were well supported, and a healthy adoption cycle followed. While all of the details of ITIL 4 are not available at the time of this writing enough has been shared to know that the new set of practices are targeted squarely at a better understanding, linkage, and realization of Business Value. In fact, the entire ITIL 4 as depicted above in Figure 4 encompasses a Value Chain model to demonstrate Business Value creation. If the adoption of ITIL 4 is embraced by the ITSM community and the model proves useful and applicable it should support better realization of Business Value and the documentation of the value. However, it is not clear that ITIL 4 in and of itself is a major advancement above earlier ITSM frameworks whether of the ITIL flavor or of other derivations. In fact, it is not clear that the framework makes the difference. It is equally likely that a stellar ITSM implementation following COBIT, ISO, or MOF could outperform an ITIL implementation simply due to the characteristics of the leadership, teamwork, funding, tooling, and support provided for the effort. Fundamentally, ITIL 4 can provide improved Business Value results and improved ITSM results but your mileage may vary.

## VII. WHAT IS TO BE DONE? A RECOMMENDED APPROACH

### A. *The view from the Business*

In discussing Business Value, it is important to turn around and look at IT from the point of view of the Business itself. An important lesson for the author was in working with a stakeholder years ago. By asking them what we can do for you in the upcoming year they replied in a slogan from a coffee mug as pictured – "don't make your problems my problems.". While this saying may seem trite in the context of a customer providing feedback this was actually rather profound. What the customer was saying was that they had their own business issues to deal with and they did not need additional problems coming inbound from IT. What they really wanted was for the platforms and applications to be Available, performant, defect free (reasonable or not), and provided with new features on schedule. For this customer that was the definition of Business Value. It is a simple definition and powerful. It can set priorities within IT and improve clarity of mission.

Another perspective on this is that for an effective IT Operations team there are few glory moments. One CTO mentioned that: *"An IT Ops team doing its job well is not noticed by anyone."*

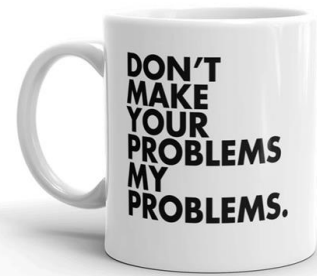

Thus, for those seeking the limelight IT Operations is probably not the place to be. Work in the ITSM space is well respected and appreciated by local management and peers but typically it is the architect and star developers who claim the garlands in creating new products and generating new revenue. Infrastructure is an assumed requirement which needs to be in place underneath those architectures and applications much like the network and the very datacenters or cloud environments powering the compute grid. Nevertheless, these are critical job roles and the better companies do place value on them just as ITSM leaders strive to demonstrate the value of ITSM to the business itself.

## B. A Simple Process

Reflecting on all the above, we can consider some core actions to take when faced with opportunities to improve processes around us and to attempt to deliver on Business Value:

1. **Observe and assess the environment**: The starting point in any process improvement activity is research, investigation, and observation. We must observe the environment and understand the intent of the business and the As-Is state of the conditions. By documenting the baseline environment through interviews, document reviews, measurements, and the like, we can build a model of the existing conditions and begin finding recommendations for improvement.

2. **Compare with best practices**: With this model in hand we compare the As-Is approach with existing best practice frameworks of relevance. There may be one or more relevant framework to compare with. This comparison will then produce clear paths for gap analysis.

3. **Develop improvement strategies**: Based on the working As-Is model and the gap analysis from the best-practices framework comparison we can develop process improvement strategies. These strategies can be comprehensive, or they can be piecemeal. They can also be targeted at particular problem areas or follow a phased approach or staged model.

4. **Conduct Pareto Analysis**: A useful technique is to apply Pareto Analysis to the improvement plans. This essentially rank orders the areas where the largest negative process impact is occurring and schedules them for the earliest resolution.

5. **Sell the approach – but don't "sell past the sell"**: In order to succeed in any process improvement effort, you must keep your stakeholders close and your questioning stakeholders even closer. Governance of the change process is vital, and communications requires constant updates on plans and progress to keep the leadership and the organization on your side – but you don't want to tax them on the effort either so keep a good balance.

6. **Execute, execute, execute**: Once you have a clear set of facts and a process design get moving and keep moving. Look for early wins and publicize them. Make some noise and let people know things are going well. You want people to be excited and get on board.

7. **The proof is in the pudding**: Theory is great. Results are a lot better. Get the process changes out there and instrument them. Have metrics in place and produce quantified and qualitative reporting regularly. As for feedback by survey and directly. Find ways to demonstrate the changes worked and then go to the next round using Pareto prioritization – there is always a next round.

## C. A Few Scenarios of Adding Business Value

Over the years applying the above process or a rough facsimile several process transformations and measurement projects have been accomplished which demonstrate true Business Value. While this paper does not allow for a recounting of all the details around these earlier efforts they have been documented and published and are available in the public domain. Here are a few such examples from the author's own portfolio of process improvement programs resulting in Business Value delivery:

- The first example covers the introduction of a software reliability engineering methodology to eliminate extensive UAT expenses and improve production software quality [CUS93]. In this case, by researching advanced test modeling methods and replacing extensive labor-intensive testing methods an 89% reduction in testing costs was realized along with other business benefits.

- Applying a Balanced Business Scorecard technique to guide the overall management of a global IT organization and focusing process improvement based on clear quantitative data reporting and analysis. Using Defects per Million to drive Network Reliability up to benefit customer quality experience directly. [CUS98]

- Developing an ITIL Incident Management based process to reduce MTTR incrementally [CUS10]. First defining a robust Incident Management process and then instrumenting the process to allow for tracking of incident start and resolution and the analysis resolution delays resulted in continued reduction in outage times then benefiting customers through shorter outages and higher Availability and directly providing higher Business Value.

## VIII.  Conclusions

Reprising the beginning of this discussion, we asked a series of questions about Business Value and ITSM. Primarily we wanted to explore if Business Value was a planned outgrowth of process engineering and process improvement efforts as delivered by ITSM efforts or was it instead a byproduct of a commonsensical and nearly instinctual quality focused ongoing responses to business, technical, and operational conditions as per the flavor of Taiichi Ohno. In other words, was Business Value a built-in requirement of the ITSM domain or was it an accidental outcome?

We first reviewed the fact that while there is no agreed to definition of Business Value there are in fact standard economic or business calculations of Business Value within finance which can be used as proxy measures for ITSM Business Value. We also outlined the emerging concepts of ITIL 4 and its central emphasis on Value Streams which in fact highlight Business Value even if in a somewhat overly abstracted and complex manner. Tying things together a simple improvement process was offered to engage on making change that can deliver true Business Value by engaging with Stakeholders on the things that matter to them and working on those issues in a top down prioritized manner and not trying to boil the ocean or get lost in a web of arcane methodology and lexicon.

In the end, you know you are adding value for the business when you are asked back to the table. This is especially true if the work you are asked to do is resulting in making things go faster, at a cheaper cost, and making them better for the customer. Of course, the trick to this is applying the right mix of process, tools, staff, skills, and metrics to get the job done. This is where the art and alchemy of the process engineer comes in. In the final analysis, Business Value is both a requirement, a mirage, and a gut instinct – a happy skilled accident as it were. Each practitioner must conjure it up for their customers the best way they can using a unique blend of creativity, insight, analysis, visioning, design, synthesis of solutions, and a drive towards continual improvement with the belief that this will all lead to an appreciable Business benefit.

## IX.  Acknowledgements

Firstly, I wish to thank Dheeraj Walia the President of the NYC itSMF Interest Group for inviting me to speak on this topic in the first place. Secondly, Dheeraj motivated me to write my ideas down in this lengthier form so there would be a more complete treatment of my somewhat offbeat angle on this topic which he believed others would appreciate. I also want to thank the other leaders of the NYC itSMF Interest Group who organized the event and continue to provide a forum for active discussion around ITSM topics. Additionally, I want to thank the members of the itSMF who participated in the talk for a lively discussion which helped advance some of my ideas on this subject and improved the content of this paper. Finally, I would like to thank Mark Smalley who took the time to provide detailed comments on this paper.

## X.  About the Author

James Cusick is an IT leader with over 30 years of experience in Software Engineering, Process Engineering, IT Operations, Cybersecurity, and Project Management. He is currently Director of IT Process Management with a global information services firm where he was also CISO & Director IT Operations. Previously James held leadership roles with Dell Services, Lucent Bell Laboratories, and AT&T Laboratories. James was Adjunct Assistant Professor of Computer Science at Columbia University and has published two recent books on IT and Software Engineering and over 75 related articles and talks. James is also a researcher in Political Economy and Innovation at the Henry George School of Social Science. Contact James at j.cusick@computer.org.

## XI.  References

1.  [BAS19] Basham, M., **ITIL Foundation: ITIL 4 Edition**, AXELOS Global Best Practice, TSO (The Stationary Office) of Williams Lea, Norwich, UK, 2019.
2.  [BAS90] Basili, V. R., "*The Goal/Question/Metric Paradigm*", White Paper, University of Maryland, 1990.
3.  [BIR10] Birk, Andreas, "*Goal/Question/Metric (GQM)*", **The Making of Software**, 11/08/2010, http://makingofsoftware.com/2010/goalquestionmetric-gqm.

4.  [BON05] Bon, J. V. **Foundations of IT Service Management: based on ITIL**, Van Haren Publishing; 2nd edition, September 15, 2005.
5.  [CUS10] Cusick, James, "Creating an ITIL Inspired Incident Management Approach: Roots, Responses, and Results", **IFIP/IEEE BDIM International Workshop on Business Driven IT Management**, Osaka, Japan, April 2010, DOI: 10.1109/NOMSW.2010.5486589.
6.  [CUS11] Cusick, J., "*A Birding Life List: A personal approach and current status"*, **A Working Paper**, June 2011, ResearchGate/Cusick/Birds.
7.  [CUS18] Cusick, James, "*Organizational Design and Change Management for IT Transformation: A Case Study"*, **Journal of Computer Science and Information Technology**, Vol. 6, No. 1, pp. 10-25, June 2018, DOI: 10.15640/jcsit.v6n1a2.
8.  [CUS19a] Cusick, James, "*Has ITSM Business Value Ever Been Spotted In The Wild*?", **itSMF USA NY Metro LIG NYC Event**, New York, NY, July 24, 2019, ResearchGate/Cusick/ITSMValue.
9.  [CUS19b] Cusick, James, J., "*A Survey of Maturity Models from Nolon to DevOps and Their Applications in Process Improvement*", arXiv:1907.01878 [cs.SE], Cornell University, July 2019.
10. [CUS98] Cusick, James, "*Reliability Surrogates for a Corporate Scorecard*", **9th Software Reliability Workshop**, Ottawa, Canada, June 1998, DOI: 10.13140/RG.2.2.32109.38884.
11. [CUS98] Cusick, James, "*Software Reliability Engineering for System Test and Production Support*", **The 4th International Conference on Applied Software Measurement**, Orlando, FL, January 1993.
12. [GAC10] Gacenga, Francis, et., al, "*An International Analysis of IT Service Management Benefits and Performance Measurement*", **Journal of Global Information Technology Management**, 13:4, 28-63, 2010, DOI: 10.1080/1097198X.2010.10856525.
13. [HER17] Hertvik, Joe, "*ITSM Frameworks: Which Are Most Popular?*", Service Management Blog, bmcblogs, April 24, 2017, https://www.bmc.com/blogs/itsm-frameworks-popular/.
14. [HUN16] Hunnebeck, Lou, "*ITIL® Practitioner: Essentials for Organizational Change Management*", **Axelos**, https://www.axelos.com/news/blogs/, 5/16/2016.
15. [itSMFa] itSMF NY Local Interest Group, https://www.itsmfusa.org/members/group.aspx?id=87662.
16. [itSMFb] itSMF USA, https://www.itsmfusa.org/default.aspx.
17. [KAP96] Kaplan, R., & Norton, D., **The Balanced Scorecard: Translating Strategy into Action**, Harvard Business Review Press; 1st edition, September 1, 1996.
18. [KEN97] Keen, Peter, **The Process Edge: Creating Value Where It Counts**, Harvard Business Review Press, 1997.
19. [MAR10] Marrone, Mauricio, et., al., Uncovering ITIL claims: IT executives' perception on benefits and Business-IT alignment, Information Systems and e-Business Management, September 2011, Volume 9, Issue 3, pp 363–380, DOI 10.1007/s10257-010-0131-7.
20. [MAS82] Massey, J., and Matsui, S, et. al., **A Field Guide to the Birds of Japan**, Kodansha International, Tokyo, 1982.
21. [OHN95] Ohno, Taiichi, **Toyota Production System: Beyond Large-scale Production**, Productivity Press Inc., 1995.
22. [PHI14] Phillip, M. A. (2014). *Delivering business value: The most important aspect of project management*", **PMI® Global Congress 2014—North America**, Phoenix, AZ, Project Management Institute, 2014.
23. [SMA18] Smalley, Mark, *Understanding How to Demonstrate the Business Value of IT*, **ITSM Tools**, September 13, 2018, https://itsm.tools/2018/09/13/understanding-how-to-demonstrate-the-business-value-of-it/.
24. [SPA18] Spacey, John, "*11 Examples of Business Value*", **Simplicable**, March 30, 2018.
25. [SUL13] Sultana, Naznin, "*A Case Study on Implementing ITIL in Business Organization - Considering Benefits with ROI*", **International Journal of Engineering Sciences & Research Technology**, ISSN 2277-9655, April 2013.
26. [WIK19] "*Business Value*", Wikipedia, the free encyclopedia, last edited on 23 July 2019, https://en.wikipedia.org/wiki/Business_value.
27. [WOM07] Womak, J., e. al., **The Machine that Changed the World**, Free Press, Reprint 2007.